\documentclass[prd,floats,floatfix,nofootinbib]{revtex4}
\usepackage[dvips]{graphicx}
\usepackage{graphics}
\usepackage{color}
\begin{document}

\title{Charged Black Holes from Interacting Vacuum}

\author{Rodrigo Maier\footnote{rodrigo.maier@uerj.br}, Manuella Corr\^ea e Silva\footnote{manuellacorrea13@gmail.com}} 

\affiliation{
Departamento de F\'isica Te\'orica, Instituto de F\'isica, Universidade do Estado do Rio de Janeiro,\\
Rua S\~ao Francisco Xavier 524, Maracan\~a,\\
CEP20550-900, Rio de Janeiro, Brazil
}

\date{\today}

\begin{abstract}
In this paper charged black holes are obtained assuming that a 
Born-Infeld electrodynamics may arise from
an interaction between the electromagnetic field and a vacuum component.
In this context Cauchy horizons do not appear in the maximal analytical extension once an event horizon is formed so that
the interior spacetime does not suffer from any sort of instabilities which are well known in the literature. On the contrary,
the causal structure exhibits an event horizon -- encapsulating a spacelike singularity -- and a cosmological horizon.
We show that the strong cosmic censorship is then restored for a wide range of the parameters including
configurations in which the black hole charge is much larger than its mass. 
We also show that the black hole thus formed described by our solution exhibits an unstable photon sphere analogous to that of the Schwarzschild metric.
%
\end{abstract}
\maketitle
\section{Introduction}
\label{intro}
Although General Relativity is the most successful theory to describe the gravitational field, some pathologies do appear when one considers the context of black hole physics. 
In fact, it is well known that all black hole classical solutions\cite{Schwarzschild:1916uq,Reissner:1916cle,nord,Kerr:1963ud,Newman:1965my} are plagued with singularities in a high energy domain so that
General Relativity must be properly corrected or even replaced in the deep UV. In the last decades considerable efforts have been made in the framework of modified theories of gravitation
in order to avoid the prediction of such singularities (see \cite{Bambi:2023try} and references therein).

While a full modified theory of gravitation remains an elusive theoretical problem, another issue which has deserved much more attention concerns the so-called Cauchy horizon. Apart from the Schwarzschild spacetime, all black hole solutions of Einstein field equations predict the formation of a Cauchy horizon $r_-$ inside the event horizon $r_+$.
In General Relativity, Cauchy horizons correspond to a global boundary in which the field equations lose their power to describe the evolution of prior initial conditions\cite{he}. Moreover, it has been shown\cite{Gursel:1979zza,chandra} that free falling observers crossing the Cauchy horizon should experience an arbitrary large blueshift of any incoming radiation so that the energy flux of test fields would diverge on it. In this sense, Cauchy horizons are known as unstable surfaces of infinite blueshift. Instabilities on Cauchy horizons have been addressed in a number of papers\cite{Poisson:1990eh,brady1,Ori:1991zz,Burko:1997zy,Burko:2002fv,Chambers:1997ef,Brady:1998au}.
A broader class of static solutions -- which emerge from a $n$-dimensional Einstein--Maxwell--scalar--$\Lambda$ system -- was also examined in \cite{Maeda:2005yd}. 
In this case it is shown that the Cauchy horizon is
unstable against kink perturbations while the event and cosmological horizons are stable.

In order to solve such puzzling issues concerning Cauchy horizons one could naively assume that General Relativity breaks down at the Cauchy horizon energy scale and a proper modified theory
-- which avoids such instabilities, for instance -- would take place. However, assuming that General Relativity does not break down at the event horizon energy scale, it is then reasonable to consider that it still holds in a given domain $r_+-\delta$ with $\delta\gtrsim0$. However, for a proper domain of parameters of a given black hole it can be shown that the Cauchy horizon can actually be arbitrarily close to the event horizon so that General Relativity could, in principle, still be applied on such energy scale. Nevertheless, as mentioned above the presence of such Cauchy horizon would still pose an insurmountable question about the stability of the spacetime inside the event horizon so that a new mechanism inside the scope of General Relativity is inexorably needed in order to avoid such spacetime instabilities.

The simplest configuration in which a Cauchy horizon appears in General Relativity is the well known Reissner-Nordstr\"om
black hole described by two parameters, namely, its charge $q$ and mass $m$. 
In this context, Einstein field equations are coupled to a linear electrodynamics
dictated by the Maxwell field. In the domain of physical interest -- in which the cosmic censorship hypothesis\cite{Penrose:1969pc} is not violated, namely, $m> 16\pi q$ --
the causal structure of the spacetime always develops an interior Cauchy horizon which makes 
the Reissner-Nordstr\"om solution ill defined. For $m\gtrsim 16\pi q$ the Cauchy horizon
may be sufficiently close to the event horizon. Therefore, assuming that General Relativity is still supposed to hold on such energy scale we are led to conjecture that a modified electrodynamics is in order to circumvent the obstacles engendered by the presence of a Cauchy horizon.

To remove the divergence of self-energy of point charges in Maxwell electrodynamics Max Born and Leopold Infeld proposed a nonlinear electrodynamics introducing a cutoff on electric fields\cite{Born:1934gh}. Years later it was noticed that Born-Infeld electrodynamics is supported by string calculations once the low effective energy action for an open superstring may lead to a Born-Infeld action\cite{Fradkin:1985qd}. It has also been shown that the Born-Infeld action may be interpreted as effective action which governs the dynamics of vector-fields on D-branes\cite{Tseytlin:1986ti}.
As a generalization of the Reissner-Nordstr\"om spacetime, black hole solutions of Born-Infeld electrodynamics are well established in the literature\cite{Dey:2004yt,Cai:2004eh}.
It has been shown\cite{Gan:2019jac} that Born-Infeld black holes may also develop Cauchy horizons but the strong cosmic censorship\cite{pen2} tends to be restored once the Born-Infeld parameter becomes small and the Cauchy horizon turns to be absent. In this sense, Born-Infeld electrodynamics seems to be a promising framework in order to avoid interior spacetime instabilities since it diminishes the parametric domain in which Cauchy horizons may appear.

In this work we give a step forward considering the assumption -- from pure phenomenological grounds -- that a Born-Infeld electrodynamics may arise from an interaction between the electromagnetic field and a vacuum component. In fact, motivated by quantum field theory considerations\cite{Lima:2013dmf,Moreno-Pulido:2020anb,Moreno-Pulido:2023ryo,SolaPeracaula:2023swx} the possibility of an interacting vacuum component has been a subject of considerable interest. From a cosmological point of view it has been shown that tensions between different observational data may be alleviated once an interacting dark energy component is considered\cite{Salvatelli:2014zta,Wang:2015wga,Zhao:2017cud,Kumar:2017dnp,Wang:2013qy,Martinelli:2019dau}.
In the context of black hole formation, it has been shown\cite{Maier:2020bgm} that a nonsingular collapse of barotropic perfect fluids may give rise to Reissner-Nordstr\"om-de Sitter black holes. Yukawa black holes, on the other hand, may also be obtained for an appropriate choice of the energy exchange between the electromagnetic field and a vacuum component\cite{Maier:2021jxv}. In the present context, we shall see that the assumption of a Born-Infeld electrodynamics arising from an interaction between the electromagnetic field and a vacuum component furnishes a solution of Einstein field equations which is free of Cauchy horizons so that the strong cosmic censorship is completely restored once an event horizon is formed.
The paper is structured as follows.
In Section \ref{sec:1} we show how a general nonlinear electrodynamics may arise from an interaction between the Maxwell field and a vacuum component. 
An exact static solution is then obtained assuming a Born-Infeld lagragian.
In Section \ref{sec:3} the causal structure is examined. 
In Section \ref{sec:4} we show that our general soltution exhibits an unstable photon sphere analogous to that of the Schwarzschild metric.
In Section \ref{sec:5} we leave our final remarks.

\section{Nonlinear electrodynamics from interacting vacuum}
\label{sec:1}
We start by considering the field equations
\begin{eqnarray}
\label{eq1}
G_{\mu\nu}-V_Ig_{\mu\nu}=-\kappa^2 T_{\mu\nu}    ,
\end{eqnarray}
where $G_{\mu\nu}$ is the Einstein tensor built with the Christoffel symbols and $\kappa^2\equiv 8\pi G$. In the above $V_I$ stands for a vacuum component which interacts with the electromagnetic field whose energy-momentum tensor reads
\begin{eqnarray}
\label{eq2}
T_{\mu\nu}=F_{\mu\alpha}F_{\nu}^{~\alpha}-\frac{1}{4}g_{\mu\nu}F.
\end{eqnarray}
Here $F$ is the Faraday scalar $F_{\alpha\beta}F^{\alpha\beta}$
with $F_{\alpha\beta} = \nabla_{\beta} A_\alpha-\nabla_\alpha A_\beta$. 

Subjecting (\ref{eq1}) to the Bianchi identities we end up with the general form
\begin{eqnarray}
\label{eq3}
\nabla_\mu T^\mu_{~\nu}=\nabla_\nu V_I=Q_\nu,
\end{eqnarray}
where we have fixed $\kappa^2\equiv 1$ for simplicity.
In the above $Q_\nu$ is a 4-vector which stands for the energy-momentum exchange between the vacuum component and the
electromagnetic field. For a vanishing $Q_\nu$ one may obtain the well known Reissner-Nordstr\"om-de Sitter solution as we should expect. In this paper, however,
we choose a different path assuming that a Born-Infeld electrodynamics may arise from such
interaction. To this end, let us consider a general nonlinear electrodynamics given by the
action
\begin{eqnarray}
S=\int\sqrt{-g}{\cal L}(F)d^4x,    
\end{eqnarray}
where ${\cal L}(F)$ is a lagrangian which sole depends on $F$. Variations of $S$ with respect to $g_{\mu\nu}$ yields
the energy-momentum tensor
\begin{eqnarray}
{\cal T}_{\mu\nu}={\cal L}_{,F} F_{\mu\alpha}F_{\nu}^{~\alpha}-\frac{1}{4}g_{\mu\nu}{\cal L},    
\end{eqnarray}
where the commas denote partial differentiation with respect to $F$. It is then easy to show that the covariant divergence 
of such energy-momentum tensor reads
\begin{eqnarray}
\nonumber
\nabla_{\mu}{\cal T}^{\mu}_{~\nu}&=&{\cal L}_{,F,F}(\nabla_\mu F)F^\mu_{~\alpha}F_\nu^{~\alpha}+{\cal L}_{,F}(\nabla_\mu F^\mu_{~\alpha})F_\nu^{~\alpha}\\
&\equiv& {\cal L}_{,F,F}(\nabla_\mu F)F^\mu_{~\alpha}F_\nu^{~\alpha}+{\cal L}_{,F}(\nabla_\mu T^\mu_{~\nu}).
\end{eqnarray}
Therefore, imposing the conservation condition $\nabla_{\mu}{\cal T}^{\mu}_{~\nu}=0$, we obtain
\begin{eqnarray}
\label{int1}
\nabla_\mu T^\mu_{~\nu}=-\frac{{\cal L}_{,F,F}}{{\cal L}_{,F}}(\nabla_\mu F)F^\mu_{~\alpha}F_\nu^{~\alpha}.   
\end{eqnarray}
Comparing the above result with (\ref{eq3}) we are then led to consider the possibility that
\begin{eqnarray}
Q_\nu=-\frac{{\cal L}_{,F,F}}{{\cal L}_{,F}}(\nabla_\mu F)F^\mu_{~\alpha}F_\nu^{~\alpha}. 
\end{eqnarray}
In this sense a general nonlinear electrodynamics may arise from an interaction between a vacuum component and 
the electromagnetic field

Let us then assume a general static spherically symmetric configuration in which the line element 
can be written as
\begin{eqnarray}
ds^2=A(r)dt^2-\frac{1}{B(r)}dr^2-r^2(d\theta^2+r^2\sin^2\theta d\phi^2).    
\end{eqnarray}
In this case the only nonvanishing component of the Faraday tensor is given by
$F_{tr}=E(r)$. From the field equations (\ref{eq1}) we then obtain $A(r)\equiv B(r)$.

From now on we shall assume that the nonlinear electrodynamics is dictated by the Born-Infeld action in which the lagrangian reads
\begin{eqnarray}
\label{BI}
{\cal L}=4\beta^2\Big(1-\sqrt{1+\frac{F}{2\beta^2}}\Big).    
\end{eqnarray}
In this case we obtain
\begin{eqnarray}
\nonumber
Q_\nu&=&\frac{1}{4\beta^2+2F}(\nabla_\mu F)F^\mu_{~\alpha}F_\nu^{~\alpha}\\
&\equiv& \frac{E^3(r)}{\beta^2-E^2(r)}\Big(\frac{dE}{dr}\Big)\delta^r_{~\nu}.
\end{eqnarray}
The substitution of (\ref{eq2}) and (\ref{BI}) in (\ref{int1}) yields the expecting result\cite{Born:1934gh}
\begin{eqnarray}
\label{el}
&&A_\mu=-\frac{q}{r}{_{2}{\cal F}_1}[1/4,1/2,5/4,-q^2/\beta^2r^4]\delta^t_{\mu},\\
\label{el12}
&&E(r)=\frac{q}{\sqrt{r^4+q^2/\beta^2}},
\end{eqnarray}
where ${_{2}{\cal F}_1}$ is the hypergeometric function and $q$ is a constant which plays a role of a
charge which regularizes the Born-Infeld static electric field.
The interacting four vector on the other hand reads
\begin{eqnarray}
\label{eq1912}
Q_\nu=-\frac{2q^4\beta^2}{r(q^2+\beta^2r^4)^2}\delta^r_{~\nu}.    
\end{eqnarray}
Here we see that $Q_\nu$ is actually well behaved in our case. In fact, since we 
have used spherical symmetry so that only $r$
dependence remains, according to (\ref{eq1912}) $Q_\nu$ is a real function of $r$
which does not diverge in the domain $r>0$. 
A straightforward integration of (\ref{eq3}) then furnishes
\begin{eqnarray}
\label{vac}
V_I=V_0-\frac{\kappa^2\beta^2}{2}\Big[\frac{q^2}{q^2+\beta^2r^4}+\ln\Big(\frac{r^4}{q^2+\beta^2r^4}\Big)\Big].    
\end{eqnarray}
From the above we notice that 
\begin{eqnarray}
\label{lambda}
\lim_{r\rightarrow \infty}V_I=V_0-\frac{\kappa^2\beta^2}{2}\Big[\ln\Big(\frac{1}{\beta^2}\Big)\Big]=:V_E ,
\end{eqnarray}
so that the effective constant $V_E$ on the RHS plays the role of an asymptotic vacuum energy, namely, a cosmological constant. In fact, it can be easily shown that for $V_E>0$ we obtain an asymptotically de Sitter spacetime. Furthermore, as $V_E\rightarrow 0$ we obtain
\begin{eqnarray}
\lim_{r\rightarrow +\infty} B(r)=1,   
\end{eqnarray}
so that the spacetime is asymptotically flat and the cosmological horizon is absent.

Finally, feeding Einstein field equations with (\ref{el})--(\ref{el12}) and (\ref{vac}) we obtain the general solution
\begin{widetext}
\begin{eqnarray}
\nonumber
&&B(r)=1-\frac{m}{8\pi r}-\frac{r^2}{3}\Big[{V_0}-\frac{\beta^2}{2}\ln\Big(\frac{r^4}{q^2+\beta^2r^4}\Big)\Big]\\
\label{ar}
&&~~~~+\frac{|q|}{3r}\sqrt{\frac{|\beta q|}{2}}\Big\{\arctan\Big({1-r\sqrt{2|\beta/ q|}}\Big)-\arctan\Big({1+r\sqrt{2|\beta/q|}}\Big)-\frac{1}{2}\ln\Big(\frac{|q|-r\sqrt{2|\beta q|}+|\beta| r^2}{|q|+r\sqrt{2|\beta q|}+|\beta| r^2}\Big)\Big\}.
\end{eqnarray}
\end{widetext}
%

\section{Causal Structure}
\label{sec:3}

From (\ref{ar}) one may easily notice that $B(r)\rightarrow -\infty$ as $r\rightarrow 0^+$.
On the other hand, a direct expansion of (\ref{ar}) furnishes
\begin{eqnarray}
\nonumber
B(r)\simeq 1-\frac{m}{8\pi r}-\frac{r^2}{3} 
\Big[V_0+\beta^2\Big(\frac{2}{3}+\frac{1}{2}\ln\Big|{\frac{q^2}{r^4}}\Big|   \Big) \Big] +\mathcal{O}(r^3)   
\end{eqnarray}
so that it can be shown that there is a domain of $\beta$
in which $B(r)$ has at most two coordinate singularities -- an event horizon $r_+$ and a cosmological horizon $r_c$ -- analogous to those of Schwarzschild-de Sitter solution. 
In this domain the strong cosmic censorship is completely restored once the Cauchy horizon is absent.
In order to numerically illustrate this feature, it is opportune to compare our solution (\ref{ar})
with the known results concerning the noninteracting case of static Born-Infeld black holes\cite{Gan:2019jac}.
In this context, (\ref{el})--(\ref{el12}) still hold but
the solution of Einstein field equations with a cosmological constant $\Lambda$ is given by
\begin{eqnarray}
\label{pure1}
ds^2=G(r)dt^2-\frac{1}{G(r)}dr^2-r^2(d\theta^2+\sin^2\theta d\phi^2),    
\end{eqnarray}
where
\begin{eqnarray}
\label{pure2}
G(r)=1-\frac{m}{8\pi r}-\frac{\Lambda r^2}{3}+\frac{2\beta^2r^2}{3}\Big(1-\sqrt{1+\frac{q^2}{\beta^2r^4}}\Big)\nonumber\\
+\frac{4q^2}{3r^2}{_{2}{\cal F}_1}[1/4,1/2,5/4,-q^2/\beta^2r^4].    \end{eqnarray}
\begin{figure}
\begin{center}
\includegraphics[width=8cm,height=5cm]{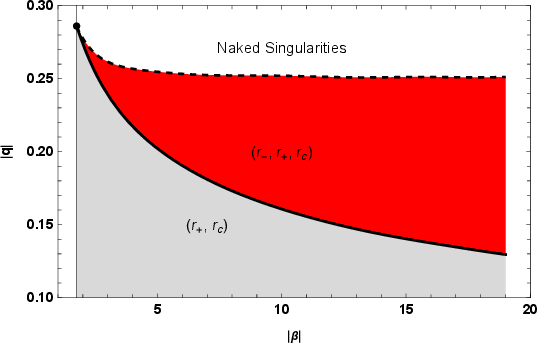}\\
\includegraphics[width=8cm,height=5cm]{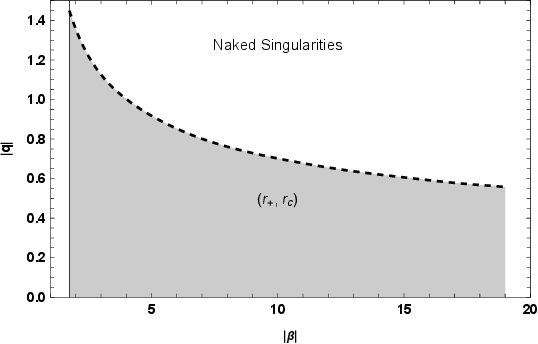}
\caption{Top panel: The parameter domain $(|\beta|, |q|)$ for static Born-Infeld black holes described by (\ref{pure1})--(\ref{pure2}). The shaded red portion is connected to a region in which
a Cauchy horizon $r_-$ appears in the maximal analytical extension. Above the dashed black curve naked singularities are obtained. Below the solid black curve Schwarzschild-de Sitter-like configurations are obtained. 
The black dot at the top left shows the minimum value of $|\beta|$ and the maximum value of $|q|$ of the red region.  
Bottom panel: the parameter domain $(|\beta|, |q|)$ described by our solution (\ref{ar}).
Above the black dashed curve naked singularities are again obtained. Below it, the an event horizon is formed and all possible configurations are free of Cauchy horizons.
In both plots we are fixing $m=4\pi$, $\Lambda=V_E=0.14$.
}
\label{fig1}
\end{center}
\end{figure}
\begin{figure}
\begin{center}
\includegraphics[width=7.5cm,height=5cm]{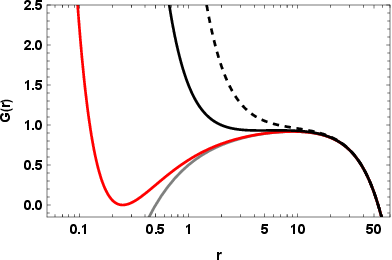}\\
\includegraphics[width=8cm,height=5cm]{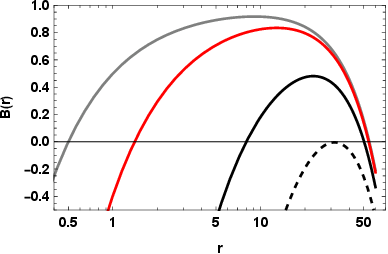}
\caption{The exact metric functions $G(r)$ (top panel) and $B(r)$ (bottom panel) given by (\ref{pure2}) and (\ref{ar}), respectively. In this plot we have fixed $m=4\pi$, $\beta=100$ and $\Lambda=V_E=10^{-3}$.
The red curves are connected to a charge parameter $q=|\bar{q}|\simeq0.25$. In the top panel we see that this corresponds to an extremal configuration while in the interacting scenario (bottom panel) an event horizon still forms.
The dashed black curves are connected to a charge parameter $q=|q_\ast|\simeq1.99$. From the bottom panel we see that this is the upper bound for the charge parameter (roughly ten times larger than ${|\bar{q}|}$) which satisfies the cosmic censorship hypothesis. Gray and black curves are connected to different charge parameters, namely $q=0$ and $q=1.00$, respectively. 
}
\label{fig2}
\end{center}
\end{figure}

\begin{figure*}
\begin{center}
\includegraphics[width=11.5cm,height=4cm]{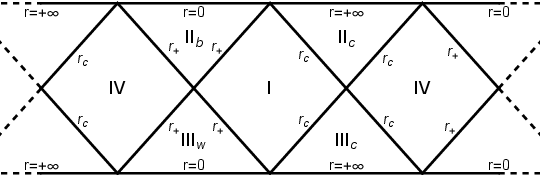}
\caption{The Penrose diagram of the solution (\ref{ar}) for the parametric domain in which an event horizon $r_+$ and a cosmological horizon $r_c$ are formed. 
In the infinite chain regions ${\rm II}_b$ denote the spacetime inside the black hole which contains the singularity at $r=0$. Beyond the event horizons $r_+$ we find the usual exterior regions ${\rm I}$ and ${\rm IV}$ bounded by the cosmological horizons $r_c$. Regions ${\rm III}_w$
are connected to classical white holes which also contain  singularities at $r=0$. Finally, regions ${\rm II}_c$ and ${\rm III}_c$ correspond to cosmological domains.
}
\label{fig3}
\end{center}
\end{figure*}
As shown in \cite{Gan:2019jac} there are two domains of physical interest in the parameter space.
In the first domain, the causal
structure develops up to three coordinate singularities -- a Cauchy horizon $r_-$, an event 
horizon $r_+$ and a cosmological horizon $r_c$. In the second domain
the Cauchy horizon is absent so that the maximal analytical extension is 
analogous to that of the Schwarzschild-de Sitter.
It is then shown that the strong cosmic censorship tends to be restored
once
the size of the first (second) domain decreases (increases)
as the Born-Infeld parameter becomes sufficiently small.  
We illustrate this behaviour in the top panel of Fig. 1 by fixing the mass as $m=4\pi$ and $\Lambda=0.14$.
The shaded red portion is connected to a region in which
a Cauchy horizon $r_-$, an event 
horizon $r_+$ and a cosmological horizon $r_c$ are formed. 
The dashed black curve corresponds to extremal black holes in which
the Cauchy horizon coincides with the event horizon. Above it 
naked singularities inside the cosmological horizon are obtained.
The solid black curve corresponds to a bifurcation in the parameter space.
Below it Schwarzschild-de Sitter-like solutions with an event horizon $r_+$
and a cosmological horizon $r_c$ are obtained.
The black dot at the top left shows the minimum value of $|\beta|$ and the maximum value of $|q|$ of the red region. In this panel we clearly see that the strong cosmic
censorship is likely to be restored as $\beta$ becomes small and the red domain decreases.
To compare the above results with our solution we consider the same range of $|\beta|$
in the paremeter domain $(|\beta|, |q|)$ in the bottom panel of Fig. 1. For $m=4\pi$ 
and $V_E=0.14$
we see a dashed black curve which divides the parameter space in two distinct regions.
The black dashed curve 
is actually connected to Nariai-like black holes\cite{Fernando:2013uza,Fernando:2013ayg} in the sense that the event horizon coincides with the cosmological horizon. 
Above it naked singularities are obtained. Below the it, 
an event horizon is formed and Schwarzschild-de Sitter-like configurations -- free of Cauchy horizons -- are obtained thus restoring the strong cosmic censorship.
%
%
%

For the last, but not least, we present another remarkable feature of our solution.
For large $\beta$ it is easy to see that (\ref{pure2})
reduces to Reissner-Nordstr\"om-de Sitter as one should expect. That is, for large $\beta$
\begin{eqnarray}
G(r)\simeq 1-\frac{m}{8\pi r}+\frac{q^2}{r^2}-\frac{\Lambda r^2}{3}.
\end{eqnarray}
Neglecting $\Lambda$ extremal black holes are obtained as long as $|\bar{q}|= m/16\pi$. 
For $|q|\leq |\bar{q}|$ the cosmic censorship hypothesis is satisfied 
once naked singularities do not appear. 
In the following we shall see that our model substantially raises the upper bound of the charge parameter which satisfies the cosmic censorship hypothesis 
in a regime of large $\beta$ and negligible $V_E$. 

For large $\beta$ it can be shown that
\begin{eqnarray}
\label{bap}
B(r)\simeq 1-\Big(\frac{m}{8\pi}+\frac{\pi\sqrt{|\beta q^3|}}{3\sqrt{2}}\Big)\frac{1}{r}-\frac{r^2}{3}V_E.  
\end{eqnarray}
Neglecting $V_E$ we see that an event horizon located at 
\begin{eqnarray}
\label{hzz}
r_+\simeq \frac{m}{8\pi}+\frac{\pi\sqrt{|\beta q^3|}}{3\sqrt{2}}    
\end{eqnarray}
is formed. Of course, the existence of $r_+$ is not guaranteed for 
arbitrarily large parameters
$q$ or $\beta$.
In fact, fixing a large $\beta$ so that $V_E$ is positive and sufficiently small a cosmological horizon will develop
as long as $|q|<|q_\ast|$, where $|q_\ast|$ is an upper limit in which the event horizon coincides with the cosmological horizon -- in the same sense of a Nariai black hole.
According to (\ref{bap}), this limit is given by
\begin{eqnarray}
\label{des24}
|q_\ast|\simeq \Big(\frac{8}{V_E|\beta|\pi^2}\Big)^{\frac{1}{3}}.    
\end{eqnarray}
In this context, a sufficient condition to obtain $|q_\ast|\gg |\bar{q}|$ is given by
\begin{eqnarray}
\label{desl}
0\ll |\beta| \ll |\bar{\beta}|:= \frac{10^5}{m^3 V_E}.    
\end{eqnarray}
As $V_E$ plays the role of a cosmological constant it must be negligible when compared to the black hole mass so that (\ref{desl})
should hold for large $|\beta|$.

To furnish a numerical illustration, let us consider $m=4\pi$ and $\Lambda=V_E=10^{-3}$. In this case we obtain $|\bar{\beta}|\simeq 5\times 10^4$
so that one may safely fix $\beta=10^2\ll|\bar{\beta}|$. In Fig. 2 we then plot the exact metric functions $G(r)$ and $B(r)$ -- given by (\ref{pure2}) and (\ref{ar}), respectively -- considering different 
charge parameters.
In this case, it can be easily show that $|\bar{q}|\simeq 0.25$ and $|q_\ast|\simeq 1.99$ -- in accordance with (\ref{des24}) -- so that the upper bound of the charge parameter to satisfy cosmic censorship hypothesis is roughly ten times larger in our model.

To proceed we now consider the analytical extension of our solution assuming the formation of an event and cosmological horizons 
analogous to that of the Schwarzschild-de Sitter solution. To start, we consider the coordinate transformations
\begin{eqnarray}
\label{t1}
\frac{2\alpha }{v}dv=\frac{1}{B(r)}dr+dt,\\
\label{t2}
\frac{2\alpha }{u}du=\frac{1}{B(r)}dr-dt.
\end{eqnarray}
In this case the metric (\ref{ar}) can be rewriten as
\begin{eqnarray}
ds^2=-\frac{4\alpha^2 B(r)}{uv}dudv-r^2(dr^2+\sin^2\theta d\phi^2).    
\end{eqnarray}
On the other hand, defining
\begin{eqnarray}
\label{tort}
r^\ast=\int \frac{1}{B(r)}dr    
\end{eqnarray}
a straightforward integration of (\ref{t1}) and (\ref{t2}) furnishes
\begin{eqnarray}
\label{i1}
r^\ast =\alpha\ln|vu|,\\
t=\alpha \ln|v/u|.    
\end{eqnarray}
From (\ref{i1}) we then obtain
\begin{eqnarray}
uv=\Big|\frac{r-r_c}{r-r_+}\Big|^{h(r)/\alpha}\sigma(r).    
\end{eqnarray}
For $r>r_+$ one may define the chart $(v_1,u_1)$ by fixing $\alpha<0$.
A straightforward integration of (\ref{tort}) shows -- with the use of (\ref{i1}) -- that $h(R)/\alpha$ and $\sigma(r)$ are positive defined functions so that the metric develops no coordinate singularity at $r_c$. However, this chart does not furnish a
regular covering for any subregion with $r<r_c$. In order to circumvent this problem, 
one may choose a different chart defined by $\alpha>0$
so that the metric develops no coordinate singularity at
$r_+$. In fact, such a chart gives a regular map of any
given subregion of the manifold with $r<r_c$. In the
domain $r_+<r<r_c$ the two charts overlap furnishing
a regular map for the entire spacetime. 
Using typical compactification functions one may easily construct the
Penrose diagram of (\ref{ar}). In Fig. 3 we illustrate it showing that it is analogous to that of the Schwarzschild-de Sitter case.

\section{Effective Photon Sphere}
\label{sec:4}

In order to examine the dynamics of test photons we now turn to
the formalism due to Carter\cite{Carter:1968rr}. To apply this procedure to 
our configuration we consider a general test particle whose the rest mass and charge are denoted by
$\bar{m}$ and $\bar{q}$, respectively. We define the canonical momentum as
\begin{eqnarray}
\label{mt}
p^\mu=\frac{dx^\mu}{d\lambda}-\bar{q}A^\mu,    
\end{eqnarray}
where the parameter $\lambda$ is defined in terms of the proper time $\tau$
as $\lambda=\tau/\bar{m}$ and $A^\mu$ is given by (\ref{el}).
The equations of motion of a charged test particle moving in a gravitational and
electromagnetic field can then be obtained from the Hamiltonian
\begin{eqnarray}
\label{ham}
H=\frac{1}{2}g_{\mu\nu}(p^\mu+\bar{q}  A^\mu)(p^\nu+\bar{q} A^\nu)=\frac{\bar{m}^2}{2}.
\end{eqnarray}
As $H$ does not depend on $t$ nor $\phi$, two fundamental constants of motion
connected to the energy and angular momentum
are directly obtained, namely
\begin{eqnarray}
p_t={\cal E}~~{\rm and}~~p_\phi=-L.    
\end{eqnarray}
A third constant of motion can be obtained by defining $S$ as the Hamilton's principal function so that
$p_r=\partial S/\partial r$ and $p_\theta=\partial S/\partial \theta$
and the Hamilton constraint (\ref{ham}) can be written as
\begin{eqnarray}
\nonumber
&&\Big(\frac{\partial S}{\partial \theta}\Big)^2+L^2\csc^2\theta\\
&&+r^2\Big\{B(r)\Big(\frac{\partial S}{\partial r}\Big)^2-\frac{({\cal E}+\bar{q}A_t)^2}{B(r)}+\bar{m}^2\Big\}=0.
\end{eqnarray}
Therefore, defining the constant of motion
\begin{eqnarray}
Q=p^2_\theta+L^2\cot^2\theta,    
\end{eqnarray}
the radial component satisfies
\begin{eqnarray}
\label{pr}
p_r=\pm\frac{\sqrt{\Psi(r)}}{B(r)}    
\end{eqnarray}
where
\begin{eqnarray}
\Psi(r)\equiv({\cal E}+\bar{q}A_t)^2-\frac{B(r)}{r^2}(L^2+Q+\bar{m}^2r^2).    
\end{eqnarray}
From (\ref{mt}), (\ref{ham}) and (\ref{pr}) we then obtain:
\begin{eqnarray}
&&\Big(\frac{dt}{d\lambda}\Big)=\frac{{\cal E}+\bar{q}A_t}{{B(r)}},\\
\label{eqr}
&&\Big(\frac{dr}{d\lambda}\Big)=\pm\sqrt{\Psi},\\
\label{eqtheta}
&&\Big(\frac{d\theta}{d\lambda}\Big)=\pm\frac{\sqrt{Q-L^2\cot^2\theta}}{r^2},\\
&&\Big(\frac{d\phi}{d\lambda}\Big)=\frac{L}{r^2}\csc^2\theta.
\end{eqnarray}

According to (\ref{eqr}) and (\ref{eqtheta}) we see that physical orbits are those
which satisfy the necessary conditions $\Psi\geq 0$ and $Q>L^2\cot^2\theta$.
To simplify our analysis we draw our attention
to circular orbits in the equatorial plane by fixing initial conditions
$\theta_0=\pi/2$ and $d\theta/d\tau|_0=0$ (which implies, $Q=0$). In this case,
circular orbits are now defined by
\begin{eqnarray}
\label{psis1}
&&\Psi|_{\tilde{r}}=0,\\
\label{psis2}
&&\frac{d\Psi}{dr}\Big|_{\tilde{r}}=0.
\end{eqnarray}
Denoting by $r_c$ the radius of circular orbits it can then be shown that a solution of (\ref{psis1})--(\ref{psis2}) is given
by
\begin{eqnarray}
\label{c1}
\frac{L}{\bar{m}}=\pm r^{3/2}\sqrt{\frac{dB/dr-2\zeta E({\cal E}+\bar{q}A_t)}{2B-r(dB/dr)}}\Big|_{r_c},
\end{eqnarray}
where $\zeta=\bar{q}/\bar{m}$ is the charge-mass ratio and
\begin{eqnarray}
\nonumber
\frac{{\cal E}}{\bar{m}}=-\zeta \Big[A_t+\frac{rEB}{2B-r(dB/dr)}   \Big]\Big|_{r_c}~~~~~~~~\\
\label{ang}
\pm\frac{B\sqrt{2(2B-rdB/dr)+(\zeta r E)^2}}{2B-r(dB/dr)}\Big|_{r_c}.
\end{eqnarray}
From (\ref{c1})--(\ref{ang}) we notice that the sufficient 
conditions (\ref{psis1})--(\ref{psis2}) for the existence of circular orbits are rather involved
once $A_t$ is given by (\ref{el}). 
To simplify our analysis we consider the case of a large $\beta$ 
black hole so that an expansion of $A_t$ furnishes
\begin{eqnarray}
\label{apr}
A_t\simeq-\frac{q}{r}+\mathcal{O}(q^3/\beta^2r^5).  
\end{eqnarray}
Taking into account (\ref{bap}) together with $V_E\rightarrow 0$, we obtain
\begin{eqnarray}
\frac{L}{\bar{m}}\simeq\pm\Big[\frac{r_+r^2+2q\zeta r(q\zeta  -r{\cal E}/\bar{m}  )}{2r-3r_+}  \Big]^{1/2}\Big|_{r_c}    ,
\end{eqnarray}
where,
\begin{eqnarray}
\label{en2}
\nonumber
\frac{{\cal E}}{\bar{m}}\simeq \frac{\zeta q r(r-2r_+)\pm r(r-r_+)\sqrt{\zeta^2q^2+2r(2r-3r_+)}}{r^2(2r-3r_+)}\Big|_{r_c}.
\end{eqnarray}
%
From the above we notice that neutral particles may develop circular orbits once
\begin{eqnarray}
(2r_c-3r_+)\geq 0.    
\end{eqnarray}
In the limiting case which equality holds we obtain an orbit with infinite energy per unit of mass, namely, a photon orbit. Therefore, in this approximation the photon sphere
is located at
\begin{eqnarray}
\label{pts}
r_{ph}\simeq\frac{3}{2}\Big(\frac{m}{8\pi}+\frac{\pi\sqrt{|\beta q^3|}}{3\sqrt{2}}\Big) ,   
\end{eqnarray}
according to (\ref{hzz}). Here we see that the second term on the RHS of (\ref{pts})
shows that the effective photon sphere is shifted away by a factor 
$\propto \sqrt{|\beta q^3|}$ from the Schwarzschild photon sphere. 

As discussed in section III, it is worth noting that the existence of $r_+$ is not guaranteed for an arbitrarily large $\beta$.
In fact, fixing the black hole mass and charge there is an upper limit $\beta=\beta_\ast$ in which the event horizon coincides with the cosmological horizon and beyond such limit naked singularities are restored.
According to (\ref{bap}), this limit is given by
\begin{eqnarray}
\label{des24}
|\beta_\ast|\simeq \frac{8}{V_E\pi^2|q|^3}.    
\end{eqnarray}
Restricting ourselves to configurations in which $|\beta|<|\beta_\ast|$ we notice 
from (\ref{pts}) that the second term on the RHS
shows that the effective photon sphere is shifted away by a factor 
$\propto \sqrt{|\beta q^3|}$ from the Schwarzschild photon sphere.
The careful reader may notice that the upper bound given in (\ref{des24}) is compatible with our approximation in (\ref{apr}). In fact, the considering the first correction term in (\ref{apr}) we obtain
\begin{eqnarray}
A_t\simeq -\frac{q}{r}+\frac{q^3}{10\beta^2 r^5}+\mathcal{O}(1/\beta^4).    
\end{eqnarray}
It can be easily shown that the second term on the RHS is negligible when compared to the first as long as
$|\beta|>>|q|/\sqrt{10} r^2$. In this case the condition $|\beta_\ast|>>|q|/\sqrt{10} r^2$ turns into
\begin{eqnarray}
V_E<<\frac{8\sqrt{10} r^2}{q^4\pi^2}.    
\end{eqnarray}
Assuming that $V_E$ plays the role of a small cosmological constant the above condition should be satisfied for typical radii/charge scales. It can be shown that this is the case when one considers the parameters used in Figs. 1 and 2.

\section{Final Remarks}
\label{sec:5}

In this paper a new route towards strong cosmic censorship hypothesis is probed assuming that
a Born-Infeld electrodynamics may arise from
an interaction between the electromagnetic field and a vacuum component.
In this context an static spherically symmetric solution of Einstein field equations
is obtained. From its maximal analytical extension we notice that 
an event horizon is formed and
the interior spacetime does not suffer from any sort of instabilities which are well known in the literature. Apart from the event horizon, the causal structure exhibits a cosmological horizon connected to the Born-Infeld parameter.
In this framework the strong cosmic censorship is then restored for a wide range of the parameters including
configurations in which the black hole charge is much larger than its mass.
The black hole thus formed described by our solution exhibits an unstable photon sphere shifted away by a factor 
$\propto \sqrt{|\beta q^3|}$ from the Schwarzschild photon sphere.

The assumption of an extension of Maxwell's theory towards a nonlinear electrodynamics -- constructed from pure phenomenological grounds -- is not a new feature in the theoretical physics mainstream.
In fact, apart from Born-Infeld theory, other models of nonlinear electrodynamics have been considered in the last decades
\cite{Kruglov:2016ezw,Kruglov:2017fck,Gullu:2020ant,Balakin:2021jby,Balakin:2021arf}. 
One of considerable interest is the so-called ModMax model\cite{Bandos:2020jsw}. It has been shown that 
ModMax nonlinear electrodynamics is the sole generalization of Maxwell's theory which
mantains conformal invariance and all other basic symmetries such as electric-magnetic duality\cite{Kosyakov:2020wxv}. As a future perspective we then intend to study the impact of different
nonlinear electrodynamics models assuming that they arise from an interacting vacuum component.

We remark that the solutions found in this paper are rather phenomenological since the energy-momentum exchange between the vacuum component and the electromagnetic field
lack of a complete and proper microphysical motivation. Nonetheless, one may regard the above results 
as an effective model to describe a new physics of black holes which restores the strong 
cosmic censorship once the Cauchy horizon is absent.
The complete quest of microphysical grounds/original mechanisms to sustain such energy-momentum exchange between the vacuum component and the electromagnetic field will be a subject of a further publication.

\section{Acknowledgments}

RM acknowledges financial support from
FAPERJ Grant No. E-$26/010.002481/2019$.
MCS acknowledges financial support of the Coordenação de Aperfeiçoamento de Pessoal de Nível Superior (CAPES) - Finance Code No. 001.



\begin{thebibliography}{99}

\bibitem{Schwarzschild:1916uq}
K.~Schwarzschild,
Sitzungsber. Preuss. Akad. Wiss. Berlin (Math. Phys. ) \textbf{1916}, 189-196 (1916).
[arXiv:physics/9905030 [physics]].

\bibitem{Reissner:1916cle}
H.~Reissner,
Annalen Phys. \textbf{355}, no.9, 106-120 (1916).

\bibitem{nord}
G.~Nordstr\"om,
Koninklijke Nederlandsche Akademie van Wetenschappen Proceedings, vol. 20, iss. 2, p.1238-1245 (1918).

\bibitem{Kerr:1963ud}
R.~P.~Kerr,
Phys. Rev. Lett. \textbf{11}, 237-238 (1963).

\bibitem{Newman:1965my}
E.~T.~Newman, R.~Couch, K.~Chinnapared, A.~Exton, A.~Prakash and R.~Torrence,
J. Math. Phys. \textbf{6}, 918-919 (1965).

\bibitem{Bambi:2023try}
C.~Bambi, {\it Regular Black Holes}, (Springer Singapore, 2023).

\bibitem{he}
S. W. Hawking and G. F. R. Ellis, {\it The large scale structure of Space-Time}, (Cambridge University Press,
Cambridge, 1973).

\bibitem{Gursel:1979zza}
Y.~Gursel, V.~D.~Sandberg, I.~D.~Novikov and A.~A.~Starobinsky,
Phys. Rev. D \textbf{19}, 413-420 (1979).

\bibitem{chandra}
S. Chandrasekhar and J. B. Hartle, 
{\it Proceedings of the Royal Society of London. Series A, Mathematical
and Physical Sciences}, 384(1787):301{315, 1982}.

\bibitem{Poisson:1990eh}
E.~Poisson and W.~Israel,
Phys. Rev. D \textbf{41}, 1796-1809 (1990).

\bibitem{brady1}
P.R. Brady, Prog. Theor. Phys. Suppl. {\bf 136}, 29 (1999).

\bibitem{Ori:1991zz}
A.~Ori,
Phys. Rev. Lett. \textbf{67}, 789-792 (1991).

\bibitem{Burko:1997zy}
L.~M.~Burko,
Phys. Rev. Lett. \textbf{79}, 4958-4961 (1997)
[arXiv:gr-qc/9710112 [gr-qc]].

\bibitem{Burko:2002fv}
L.~M.~Burko,
Phys. Rev. Lett. \textbf{90}, 121101 (2003)
[erratum: Phys. Rev. Lett. \textbf{90}, 249902 (2003)]
[arXiv:gr-qc/0209084 [gr-qc]].

\bibitem{Chambers:1997ef}
C.~M.~Chambers,
Annals Israel Phys. Soc. \textbf{13}, 33 (1997)
[arXiv:gr-qc/9709025 [gr-qc]].

\bibitem{Brady:1998au}
P.~R.~Brady, I.~G.~Moss and R.~C.~Myers,
Phys. Rev. Lett. \textbf{80}, 3432-3435 (1998)
[arXiv:gr-qc/9801032 [gr-qc]].

\bibitem{Maeda:2005yd}
H.~Maeda, T.~Torii and T.~Harada,
Phys. Rev. D \textbf{71}, 064015 (2005)
[arXiv:gr-qc/0501042 [gr-qc]].

\bibitem{Penrose:1969pc}
R.~Penrose,
Riv. Nuovo Cim. \textbf{1}, 252-276 (1969).

\bibitem{Born:1934gh}
M.~Born and L.~Infeld,
Proc. Roy. Soc. Lond. A \textbf{144}, no.852, 425-451 (1934).

\bibitem{Fradkin:1985qd}
E.~S.~Fradkin and A.~A.~Tseytlin,
Phys. Lett. B \textbf{163}, 123-130 (1985).

\bibitem{Tseytlin:1986ti}
A.~A.~Tseytlin,
Nucl. Phys. B \textbf{276}, 391 (1986)
[erratum: Nucl. Phys. B \textbf{291}, 876 (1987)].

\bibitem{Dey:2004yt}
T.~K.~Dey,
Phys. Lett. B \textbf{595}, 484-490 (2004)
[arXiv:hep-th/0406169 [hep-th]].

\bibitem{Cai:2004eh}
R.~G.~Cai, D.~W.~Pang and A.~Wang,
Phys. Rev. D \textbf{70}, 124034 (2004)
[arXiv:hep-th/0410158 [hep-th]].

\bibitem{Gan:2019jac}
Q.~Gan, G.~Guo, P.~Wang and H.~Wu,
Phys. Rev. D \textbf{100}, no.12, 124009 (2019).
[arXiv:1907.04466 [hep-th]].

\bibitem{pen2} R. Penrose, {\it Singularities of Spacetime}, Theoretical Principles in Astrophysics and Relativity (A78-43851
19-90), Chicago University Press, Chicago (1978).

\bibitem{Lima:2013dmf}
J.~A.~S.~Lima, S.~Basilakos and J.~Sola,
Mon. Not. Roy. Astron. Soc. \textbf{431}, 923-929 (2013)
[arXiv:1209.2802 [gr-qc]].

\bibitem{Moreno-Pulido:2020anb}
C.~Moreno-Pulido and J.~Sola,
Eur. Phys. J. C \textbf{80}, no.8, 692 (2020)
doi:10.1140/epjc/s10052-020-8238-6
[arXiv:2005.03164 [gr-qc]].

\bibitem{Moreno-Pulido:2023ryo}
C.~Moreno-Pulido, J.~Sola Peracaula and S.~Cheraghchi,
Eur. Phys. J. C \textbf{83}, no.7, 637 (2023)
[arXiv:2301.05205 [gr-qc]].

\bibitem{SolaPeracaula:2023swx}
J.~Sola Peracaula, A.~Gomez-Valent, J.~de Cruz Perez and C.~Moreno-Pulido,
Universe \textbf{9}, no.6, 262 (2023)
[arXiv:2304.11157 [astro-ph.CO]].

\bibitem{Salvatelli:2014zta}
V.~Salvatelli, N.~Said, M.~Bruni, A.~Melchiorri and D.~Wands,
Phys. Rev. Lett. \textbf{113}, no.18, 181301 (2014)
[arXiv:1406.7297 [astro-ph.CO]].

\bibitem{Wang:2015wga}
Y.~Wang, G.~B.~Zhao, D.~Wands, L.~Pogosian and R.~G.~Crittenden,
Phys. Rev. D \textbf{92}, 103005 (2015)
[arXiv:1505.01373 [astro-ph.CO]].

\bibitem{Zhao:2017cud}
G.~B.~Zhao, M.~Raveri, L.~Pogosian, Y.~Wang, R.~G.~Crittenden, W.~J.~Handley, W.~J.~Percival, F.~Beutler, J.~Brinkmann and C.~H.~Chuang, \textit{et al.}
Nature Astron. \textbf{1}, no.9, 627-632 (2017)
[arXiv:1701.08165 [astro-ph.CO]].

\bibitem{Kumar:2017dnp}
S.~Kumar and R.~C.~Nunes,
Phys. Rev. D \textbf{96}, no.10, 103511 (2017)
[arXiv:1702.02143 [astro-ph.CO]].


\bibitem{Wang:2013qy}
Y.~Wang, D.~Wands, L.~Xu, J.~De-Santiago and A.~Hojjati,
Phys. Rev. D \textbf{87}, no.8, 083503 (2013)
[arXiv:1301.5315 [astro-ph.CO]].

\bibitem{Martinelli:2019dau}
M.~Martinelli, N.~B.~Hogg, S.~Peirone, M.~Bruni and D.~Wands,
Mon. Not. Roy. Astron. Soc. \textbf{488}, no.3, 3423-3438 (2019)
[arXiv:1902.10694 [astro-ph.CO]].

\bibitem{Maier:2020bgm}
R.~Maier,
Int. J. Mod. Phys. D \textbf{29}, no.14, 2043023 (2020)
[arXiv:2005.09576 [gr-qc]].



%

\bibitem{Maier:2021jxv}
R.~Maier,
Class. Quant. Grav. \textbf{39}, 155008 (2022)
[arXiv:2108.04911 [gr-qc]].

\bibitem{Fernando:2013uza}
S.~Fernando,
Int. J. Mod. Phys. D \textbf{22}, no.13, 1350080 (2013)
[arXiv:1312.7801 [gr-qc]].

\bibitem{Fernando:2013ayg}
S.~Fernando,
Gen. Rel. Grav. \textbf{45}, 2053-2073 (2013)
[arXiv:1401.0714 [gr-qc]].

\bibitem{Carter:1968rr}
B.~Carter,
Phys. Rev. \textbf{174}, 1559-1571 (1968).


\bibitem{Kruglov:2016ezw}
S.~I.~Kruglov,
Annalen Phys. \textbf{528}, 588-596 (2016)
[arXiv:1607.07726 [gr-qc]].

\bibitem{Kruglov:2017fck}
S.~I.~Kruglov,
Annals Phys. \textbf{378}, 59-70 (2017)
[arXiv:1703.02029 [gr-qc]].

\bibitem{Gullu:2020ant}
I.~Gullu and S.~H.~Mazharimousavi,
Phys. Scripta \textbf{96}, no.4, 045217 (2021)
[arXiv:2009.08665 [gr-qc]].

\bibitem{Balakin:2021jby}
A.~B.~Balakin and A.~A.~Galimova,
Phys. Rev. D \textbf{104}, no.4, 044059 (2021)
[arXiv:2106.01417 [gr-qc]].

\bibitem{Balakin:2021arf}
A.~B.~Balakin, V.~V.~Bochkarev and A.~F.~Nizamieva,
Symmetry \textbf{2021}, 13 (2038)
[arXiv:2110.03005 [gr-qc]].


\bibitem{Bandos:2020jsw}
I.~Bandos, K.~Lechner, D.~Sorokin and P.~K.~Townsend,
Phys. Rev. D \textbf{102}, 121703 (2020)
[arXiv:2007.09092 [hep-th]].

\bibitem{Kosyakov:2020wxv}
B.~P.~Kosyakov,
Phys. Lett. B \textbf{810}, 135840 (2020)
[arXiv:2007.13878 [hep-th]].


\end{thebibliography}
\end{document}